\documentclass[prd,nofootinbib,showpacs,preprint]{revtex4}

\usepackage{graphicx}
\usepackage{amsmath,amssymb}
\usepackage{array}

\usepackage[colorlinks]{hyperref}
\usepackage[usenames,dvipsnames]{color}

\newcommand{\vev}[1]{\langle #1 \rangle}
\newcommand{\fcal}{\mathcal{F}}

\begin{document}

\title{Spontaneously broken parity and consistent cosmology with transitory domain walls}
\author{Sasmita Mishra}
\email{sasmita@phy.iitb.ac.in}
\author{Urjit A. Yajnik}
\email{yajnik@iitb.ac.in}
\affiliation{Indian Institute of Technology, Bombay, Mumbai 400076, India
and Indian Institute of Technology, Gandhinagar, Ahmedabad 382424, India}
\date{}

\begin{abstract}
Domain wall structure which may form in theories with spontaneously broken 
parity is generically in conflict with standard cosmology. 
It has been argued that Planck scale suppressed effects can be sufficient 
for removing such domain walls. We study this possibility for three specific
evolution scenarios for the domain walls, with evolution during radiation dominated
era, during matter dominated era, and that accompanied by weak inflation.
We determine the operators permitted by the supergravity formalism
and find that the field content introduced to achieve desired spontaneous parity 
breaking makes possible Planck scale suppressed terms which can potentially remove 
the domain walls safely. However, the parity breaking scale, equivalently 
the majorana mass scale $M_R$ of the right handed neutrino, does get constrained
in some of the cases, notably for the matter dominated evolution case which 
would be  generic to string theory inspired models giving rise to moduli fields.
One left-right symmetric model with only triplets and bidoublets is found 
to be more constrainted than another admitting a  gauge singlet.

\end{abstract}
\pacs{12.10.-g,12.60.Jv}
\maketitle
 
\section{Introduction}
\label{sec:intro}

Existence of right handed neutrino states \cite{Fukuda:2001nk,Bahcall:2004mz,
Ahmad:2002ka,Ahmad:2002jz} is a strong indicator of parity
as a fundamental symmetry of nature, spontaneously broken in low energy physics.
The scale of parity breakdown is as yet unknown. The see-saw mechanism
\cite{Minkowski:1977sc, Gell-Mann:1980vs, Yanagida:1979as, Mohapatra:1979ia} while
providing elegant qualitative explanation is unable to make a narrow prediction
of the relevant energy scale due to wide variation in the fermion masses across the
generations. In this paper we study the role of domain walls formed by
spontaneous parity breaking in determining or constraining the energy
scale of breakdown of this symmetry.

To be specific we study the implementation of parity in Left-Right symmetric
models\cite{Pati:1974yy, Mohapatra:1974gc, Senjanovic:1975rk, Mohapatra:1980qe, Deshpande:1990ip}. While such models can descend from an $SO(10)$ grand unified theory
\cite{Kuchimanchi:1993jg,Kuchimanchi:1995vk,Aulakh:1997fq,Aulakh:1998nn,Aulakh:1997ba},
the scale of such unification is known to
be high($\sim10^{16} {\rm GeV}$).
On the other hand the scale of parity breaking is completely undetermined
and could be much lower. Indeed there are no observational 
obstructions to the  parity breaking scale, and the associated right 
handed neutrino majorana  mass scale, from being as low as 
TeV scale \cite{Lindner:1996tf,Raidal:2000ru,Sahu:2004sb}.
Here we study a particular source of constraint on this scale 
imposed by cosmology, with the possibility of restricting the parity
breaking scale to low values. It suffices for this purpose to focus on 
the Left-Right symmetric model alone, independent of how the model may
unify into $SO(10)$. For the purpose of protecting the low scale theory 
from large radiative corrections we impose supersymmetry.
Specifically we investigate the constraints placed on models incorporating 
TeV scale supersymmetry.

A robust consequence of approximate left-right symmetry in the early universe
is the occurance of transient domain walls. It has been proposed 
\cite{Rai:1992xw} that such domain walls are susceptible  to instability
arising from non-renormalizable operators suppressed by Planck scale.
Supergravity then introduces two interesting ingredients not considered
in the previous treatments of domain walls.  Firstly the structure of
the non-renormalizable terms is dictated by the supersymmetry
formalism\cite{Martin:1997ns,Chung:2003fi,Weinberg:2000cr}.
On the other hand one has to contend with the danger of gravitino
overabundance\cite{Ellis:1984er,Asaka:2000zh}. 
In this paper we explore the restrictions on the
possible energy scale of parity breaking imposed by these considerations.

We study these effects in the context of two implementations of left-right
symmetry, one where all superfields carry non-trivial
gauge couplings and another, for comparison, which contains a gauge singlet.  
We also study these models within three different scenarios for the
dynamics of the wall complex. One is a scenario in which the walls 
disappear within the radiation dominated era, another where dominance of
moduli keeps the universe matter dominated during
the domain wall evolution, and a third wherein the domain walls
in fact come to dominate the universe for a limited epoch, accompanied
by a mild inflationary phase.
In all the scenarios of domain wall evolution, the left-right symmetric 
model with a singlet 
turns out to be less restricted than the one without singlets.
The model without any singlet turns out, at least in one scenario of
wall evolution, to be sufficiently restrictive that the parity breaking scale
can be no larger than $10^8$GeV.
The overall lesson is that the new features introduced
by supergravity can have a strong bearing on the scale of parity breaking
for ensuring viable cosmology free of permanent domain walls.

The alternative to supergravity induced terms for distinguishing between
parity symmetric vacua was studied in \cite{Yajnik:2006kc,Mishra:2008be}
wherein the parity breaking operators are
induced at a much lower scale, viz., the supersymmetry breaking scale, and
signalled by the gauge mediation mechanism, thus linking the two scales.
This avenue for evading unwanted domain walls remains open for models 
that get restricted in the present study.

In the remainder of the paper, in sec. \ref{sec:dicsymmcosmo} we review 
cosmology with domain walls, in sec. \ref{sec:modelsofdynamics} we discuss
three possible scenarios for evolution of domain walls and the requirement
in each case for the successful destabilization of the wall complex. 
In sec. \ref{sec:sugra} we discuss the origin of the parity breaking
terms within supergravity formalism. In sec. \ref{sec:lrsm} we discuss 
the essentials of supersymmetric left-right model.
In subsection \ref{subsec:ABMRSmodel} we discuss a particular renormalisable
implementation of left-right supersymmetric model and then check for the
sufficiency of the induced Planck scale terms to cause the required 
destabilisation of the domain walls.
The same study is carried out for a recent implementation of supersymmetric 
left-right symmetry including a gauge singlet in subsection 
\ref{subsec:BMmodel}. In sec. \ref{sec:concl} we summarise the conclusions.
Two appendices, appendix \ref{sec:ap1} and appendix \ref{sec:ap2} contain 
the calculations of  the Planck scale suppressed terms for the two left-right 
symmetric models studied.

\section{Discrete symmetry breaking and cosmology}
\label{sec:dicsymmcosmo}
Spontaneous breakdown of discrete symmetries in unified models can 
give rise to domain walls. Stable domain walls from unified theories 
have long been recognized as  inconsistent with the observed Universe\cite{Kibble:1980mv}.
In the presence of several degenerate ground
states the domain wall network in the Universe can be rather complicated.
If the domain walls are stable, the energy density 
stored in the network decreases as $\rho \propto 1/a$, resulting
in $a(t) \propto t^2$ leading to a mild inflationary behavior.
A more generic possibility is that the walls continue to be destroyed
due to collisions and result in the formation of homogeneous domains. 
However even one domain wall of grand unified scale of the size of the horizon 
can conflict with cosmology. For domain walls arising at symmetry breaking
scale $M_R$, it can be estimated \cite{Zeldovich:1974uw,Kibble:1980mv} that
the density perturbation introduced by them would conflict with known
magnitude of temperature fluctuation $\Delta T/T \sim 10^{-5}$ of the 
cosmic microwave background if $M_R \gtrsim 1$MeV. 
This impasse is overcome if the spontaneously broken discrete symmetry is
also broken explicitly by a small amount. For example, \cite{Vilenkin:1984ib}
the symmetry $\phi \rightarrow -\phi$ can be broken by adding a term
$\epsilon\phi^3$ to the Lagrangian  which 
gives a pressure difference governed by the small parameter $\epsilon$
between the two sides of the domain walls.

The authors of \cite{Rai:1992xw} 
have  discussed several similar reasons for considering such gravity induced 
terms and their effect in destabilizing domain walls.
For the theory of a generic neutral scalar field $\phi$, the effective higher 
dimensional operators can be written as 
\begin{equation} 
V_{eff} = \frac{C_5}{M_{Pl}} \phi^5  + \frac{C_6}{M_{Pl}^2}\phi^6 +...  
\label{eq:Veff} 
\end{equation}
Such terms give rise to a pressure difference across a given domain wall
of the amount of the difference in the effective energy density across the
wall, $\delta\rho=\Delta V_{eff}$. Specifically, the terms odd in $\phi$ break 
the discrete symmetry, and the leading contribution to the difference in 
pressure is $\frac{2 C_5}{M_{Pl}} \langle \phi \rangle^5$.
From cosmological considerations we can separately estimate the difference 
$\delta\rho$ in the energy density across a domain wall required for  
timely removal of the  domain walls. It is found that this has a value 
smaller by a factor $M_R/M_{Pl}$ than the leading order term in the generic 
effective potential considered above, leading to the conclusion that 
the walls will indeed be removed  without conflicting with cosmology.

This toy example however is only instructional because in realistic
theories, gauge invariance and supersymmetry significantly constrain the
structure of terms that can arise. In a non-supersymmetric 
example \cite{Lew:1993yt}, gauge invariance implies that the leading 
order operator is suppressed by $1/M_{Pl}^2$. Further, the terms 
are products of vacuum expectation values of different scalar fields which 
may differ significantly in their mass scales, as will be true in our
study. The exceptional case
where the toy example may be of direct relevance is the presence of
one or more gauge singlet scalar fields in the theory, ( equivalently,
superfields in a supersymmetric theory) permitting the kind of terms listed above. 

As an illustration of this phenomenon consider the leading order operator 
containing several scalar fields,
$\phi_i$
\begin{equation}
 \frac{\phi_1 \phi_2 \phi_3\phi_4\phi_5}{M_{Pl}}
\label{eq:5phi}
\end{equation}
Borrowing from a calculation that will be detailed later (see eq. (\ref{eq:eps-v11half}) ), 
suppose the
required constraint for successful removal domain walls is 
$\delta\rho\gtrsim M_R^{11/2}/M_{Pl}^{3/2}$. Then the requirement that the
operator in Eq.(\ref{eq:5phi}) is sufficient for removing domain walls
is that
\begin{equation}
 \frac{\phi_1 \phi_2 \phi_3\phi_4\phi_5}{M_{Pl}} \gtrsim \frac{M_R^{11/2}}{M_{Pl}^{3/2}}
\label{eq:5phi-11half}
\end{equation}
Now suppose that there are only two kinds of fields, one getting the vacuum
expectation value of the order of $M_R$, the parity breaking scale, and 
that there are $x$ factors of this field in the operator, while the other
field constitutes the remaining factors, and gets a TeV scale value $v$. Then 
\begin{equation}
 v^{5-x}\gtrsim \frac{M_R^{{\frac{11}{2}}-x}}{M_{Pl}^{1/2}}
\end{equation}
so that, taking TeV scale to be $ v \sim 10^3$GeV,
\begin{equation}
 {\textrm log}\, M_R \lesssim \frac{24.5-3x}{5.5-x}
\label{eq:logMR}
\end{equation}
This relation means that if $x=5$, $M_R$ can be as large as the Planck scale,
while for $x=1$, $M_R $ is forced to have value $<10^5$GeV,

In our analysis we shall be assuming that parity breaking occurs at the same
mass scale as the mass of the heavy majorana neutrinos. The constraints to be 
derived also depend on a few ancillary details, specifically the dynamics of
the walls before they disintegrate. The primary implication of these other 
details, to be discussed in the following, is only the value of the
temprature at which the standard cosmology resumes. 
In the following  we shall admit the possibility of this more general 
kind of evolution and
focus attention on two issues. The Universe should be radiation dominated
at temperature $10$MeV and lower in order to ensure successful Big Bang 
Nucleosynthesis(BBN). Secondly, the danger of gravitino overabundance
is generic to all supersymmetric models. Detailed calculations  
\cite{Ellis:1984eq,Asaka:2000zh} show that the gravitinos with unacceptable 
consequences to observable cosmology are  generated entirely after reheating 
of the universe subsequent to primordial inflation provided
$T_R \gtrsim 10^9 \textrm{GeV}$.
Thus we make the requirement that entropy generated from the decay of 
domain walls should not raise the temprature of the Universe above
this temperature scale.

\section{Models of dynamics of domain wall complex}
\label{sec:modelsofdynamics}
Occurance of domain walls per se at some epoch in the
early Universe is not ruled out, provided the walls eventually get destroyed. 
Safe disappearance of domain walls 
was dealt with in some generality in \cite{Rai:1992xw} and \cite{Preskill:1991kd}, 
the former in the context of Planck scale effects, and the latter
in the context of instanton induced effects from QCD. There have been several 
model specific studies of the fate of domain walls, e.g., 
\cite{Matsuda:2000mb,Kawasaki:2004rx,Sarkar:2007ic} and 
studies pointing 
out that transient domain walls may in fact form the basis for 
explanation of some of the cosmological effects such as 
leptogenesis, \cite{BenMenahem:1992sg,Lew:1993yt,
Cline:2002ia,Sarkar:2007ic}
or address problems such as proliferation of relics\cite{Kawasaki:2004rx,
Yajnik:2006kc,Sarkar:2007ic}.

For the purpose of this paper we consider three possible routes through
which domain walls may evolve. The first one consists of domain walls originating
in radiation dominated universe and destabilized and destroyed also within
the radiation dominated era before they begin to dominate the energy
density of the Universe, referred to as RD model. The second scenario was
essentially proposed in \cite{Kawasaki:2004rx}, which consists of the walls
originating  in a radiation dominated phase, subsequent to which the 
universe enters a ``matter dominated'' phase, either due to substantial 
production of heavy unwanted relics such as moduli, or due to the presence of
a coherently 
oscillating scalar field. This we refer to as MD model.
Here too walls disappear before they come to dominate the energy density
of the Universe.
Finally, we consider a variant of the MD model in which the domain 
walls come to dominate the energy density of the Universe and continue
to do so for a considerable epoch, leading to mild inflationary behaviour
or weak inflation \cite{Lyth:1995hj},\cite{Lyth:1995ka}. We refer to it 
as WI model. We now describe these in detail.

\subsection{Evolution in radiation dominated universe}
The essentials of this scenario are as originally proposed by Kibble
\cite{Kibble:1980mv} and Vilenkin \cite{Vilenkin:1984ib}. Domain walls 
arise at some temperature 
$T_c$, the critical temperature of a phase transition at which a scalar field 
$\phi$ acquires a non-zero vacuum expectation value at a scale $M_R$. The energy 
density trapped per unit area of the wall is $\sigma \sim M_R^3$. 
The dynamics of the walls is determined by two quantities, 
force due to tension $f_T $ and force 
due to friction $f_F $. The first of these is determined by intrinsic
energy per unit area $\sigma$, and the average scale of 
radius of curvature $R$ prevailing in the wall complex.
We estimate $f_T \sim \sigma/R$. The frictional force is proportional
to the collisions encountered by the wall with surrounding radiation
with energy density $\sim T^4$, while the former is navigating through
the medium at speed $\beta$. This force is estimated as $f_F \sim \beta T^4$.
The epoch at which these two forces balance each other sets the time 
scale  $t_R \sim R/{\beta}$. We may take this as the time scale by which 
the wall portions that  started with radius of curvature scale $R$ 
straighten out. Putting together these statements we get the following 
scaling law for the growth of the scale $R(t)$ on which the wall 
complex is smoothed out.  
\begin{equation}
 R(t)\approx (G \sigma)^{1/2} t^{3/2}
\end{equation}
Now the energy density of the domain walls goes as 
$\rho_W \sim (\sigma R^2/R^3) \sim (\sigma/G t^3)^{1/2}$. 
In radiation dominated era this
$\rho_W$ becomes comparable to the energy density of the Universe
$(\rho \sim 1/(G t^2))$ around time $t_0 \sim 1/(G \sigma)$.

Next, we consider destabilization of walls due to pressure difference
$\delta \rho$ arising from small asymmetry in the conditions on the two
sides. This effect competes with the two quantities mentioned above. Since
$f_F\sim 1/(Gt^2)$ and $f_T \sim ( \sigma /(G t^3))^{1/2}$, 
it is clear that at some point of time, $\delta \rho$ would exceed
either the force due to tension or the force due to friction.  For either
of these requirements to be satisfied before $t_0 \sim 1/(G \sigma)$
we get 
\begin{equation}
 \delta \rho \ge G \sigma^2 \approx \frac{M_R^6}{M_{Pl}^2}
\sim M_R^4 \frac{M_R^2}{M_{Pl}^2}
\label{eq:eps-vsix}
\end{equation}
We may read this formula by defining a factor
\begin{equation}
 \fcal \equiv \frac{\delta \rho}{M_R^4}
\label{eq:fcal}
\end{equation}
where $M_R^4$ is the energy density in the wall complex immediately at the phase
transition, which relaxes by factor $\fcal$ at the epoch of their decay.
The factor $\fcal$ is strongly dependent on the assumed model of evolution 
of the wall complex, and is found to be $M_R^2/M_{Pl}^2$ in this model.

\subsection{Evolution in matter dominated universe}
We next take up the model of evolution in which the scale factor behaves
as in a matter dominated era by the time the domain walls get destabilised. 
This possibility was considered in \cite{Kawasaki:2004rx} by Kawasaki and Takahashi.
The analysis begins by assuming that the initially formed wall complex 
in a phase tansition is expected to rapidly relax to a few walls per horizon 
volume at an epoch characterized by Hubble parameter value $H_i$. 
Thus the initial energy density in the wall complex is 
$\rho^{(\textstyle{in})}_W \sim \sigma H_i$. This epoch onwards, we assume 
the energy density of the Universe to be dominated by heavy relics  or 
an oscillating modulus field, in either of which cases, the scale factor 
$a(t)\propto t^{2/3}$. The energy density for both of these cases scales 
as $\rho_{mod} \sim \rho^{(\textstyle{in})}_{mod}/a(t)^3$. 
If the wall complex remains frustrated, i.e. its energy density
contribution $\rho_{DW} \propto 1/a(t)$, it can be seen that
\cite{Kawasaki:2004rx}, the Hubble parameter at the epoch of
equality of DW contribution with contribution of the rest of the
matter is given by 
\begin{equation}
H_{eq} \sim \sigma^{\frac{3}{4}} H_i^{\frac{1}{4}} M_{Pl}^{-\frac{3}{2}} ~,
\label{eq:Heq}
\end{equation}
To proceed we assume that the domain walls start decaying as soon as
they dominate the energy density of the Universe. If the temperature
at this particular epoch is $T_D$, then $H_{eq}^2 \sim G T_D^4$.
So from Eq.(\ref{eq:Heq}) we find 
\begin{equation}
 T_D^4 \sim \sigma^{\frac{3}{2}}H_i^{\frac{1}{2}}M_{Pl}^{-1}
\label{eq:TD4}
\end{equation}

Let us assume the temperature at which the domain walls are formed
$ T \sim \sigma ^{1/3}$. So
\begin{equation}
 H_i^2 = \frac{8\pi}{3}G\sigma^{\frac{4}{3}} \sim 
\frac{\sigma^{\frac{4}{3}}}{M_{Pl}^2}
\label{eq:Hi}
\end{equation}
From Eq.(\ref{eq:TD4}) we get,
\begin{equation}
 T_D^4 \sim \frac{\sigma^{11/6}}{M_{Pl}^{3/2}} \sim 
\frac{M_R^{11/2}}{M_{Pl}^{3/2}} \sim M_R^4
\left( \frac{M_R}{M_{Pl}} \right)^{3/2}
\label{eq:TD4I}
\end{equation}
Now requiring $\delta\rho > T_D^4$ we get,
\begin{equation}
\delta\rho> M_R^4 \left(\frac{M_R}{M_{Pl}}\right)^{3/2}
\label{eq:eps-v11half}
\end{equation}
Thus in this case we find $\fcal \equiv (M_R/M_{Pl})^{3/2}$, a milder suppression
factor than in the radiation dominated case above.. 

\subsection{Evolution including weak inflation}
The third possibility we consider is that the walls do not disintegrate by
the time they come to dominate the energy density of the Universe, but 
in fact go on to dominate the energy density of the Universe. 
This domination however lasts for a
limited epoch. Since the universe evolves as $a(t) \propto t^2$, it
leads to an epoch of mild inflation (as against exponential) also 
referred to as thermal or weak inflation.
This possibility has been considered \cite{Lyth:1995hj},\cite{Lyth:1995ka} 
in the context of removal moduli in superstring cosmology \cite{Coughlan:1983ci,Banks:1993en,
deCarlos:1993jw}.
Such a situation is most likely in the case when the
$\delta \rho$ is typically small, not large enough
to destabilize the walls sufficiently quickly. But eventually a small
pressure difference will also win
over $f_T$ or $f_F$, because either the curvature scale $R$ diverges,
as for straightened out walls, or the translational speed $s$ reduces
drastically. Since we have no microscopic model for deciding which
of these is finally responsible, we introduce a temperature scale
$T_D$ at which the walls begin to experience instability.
Note that unlike in the previous example, we will not be able to
estimate $T_D$ in terms of other mass scales and will accept it as 
undetermined and consider a few reasonable values for it for our final
estimate.


As has been studied above, at $H_{eq}$ the energy density of the domain
wall network dominates energy density of the Universe.
The scale factor at this  epoch is characterized by
$a_{eq}$. Denoting the energy density of the domain walls
at the time of equality as $\rho_{DW}(t_{eq})$,
the evolution of energy density can be written as,
\begin{equation}
 \rho_{DW}(t_d)\sim \rho_{DW}(t_{eq})\left( \frac{a_{eq}}{a_d}\right)
\label{eq:rhodw}
\end{equation}
where $a_d$ is scale factor at the epoch of decay of domain
wall corresponding to time $t_d$. If the domain walls decay at an
epoch characterized by temperature $T_D$, then $\rho_{DW}(t_d)\sim T_D^4$.
So from Eq.(\ref{eq:rhodw}),
\begin{equation}
 T_D^4=
 \rho_{DW}(t_{eq})\left( \frac{a_{eq}}{a_d}\right)
\label{eq:rhoeq}
\end{equation}

In the matter dominated era the energy density of the moduli
fields scale as,
\begin{equation}
 \rho^d_{mod} \sim \rho^{eq}_{mod}\left( \frac{a_{eq}}{a_d}\right)^3
\label{eq:rhoMod}
\end{equation}
Substituting the value of $a_{eq}/a_d$ from Eq.(\ref{eq:rhoeq}) in
the above equation,
\begin{equation}
 \rho^d_{mod} \sim \frac{T_{D}^{12}}{\rho^2_{DW}(t_{eq})}
\label{eq:rhoModTD}
\end{equation}
Since the energy density of the domain walls dominates
the universe after the time of equality,
$\rho_{DW}(t_d) > \rho^d_{mod}$. So the pressure
difference across the domain walls when they start
decaying is given by,
\begin{equation}
 \delta \rho \gtrsim \frac{T_D^{12} G^2}{H_{eq}^4}
\label{eq:deltaRho}
\end{equation}
where we have used the relation $H_{eq}^2 \sim G \rho_{DW}(t_{eq})$.
Replacing the value of $H_{eq}$ from Eq.(\ref{eq:Heq}),  and 
$H_i^2 \sim G \rho^{in}_{DW}\sim M_R^4/M_{Pl}^2$, 
\begin{equation}
 \delta \rho \gtrsim M_R^4 \left(\frac{T_D^{12} M_{Pl}^3}
{M_R^{15}}\right)
\label{eq:deltaRhoTD}
\end{equation}
The $\fcal$ factor introduced in eq. (\ref{eq:fcal}) turns out in this case to
be $(T_D^{12} M_{Pl}^3)/{M_R^{15}}$, rather sensitively dependent upon $T_D$.

\section{Supergravity and left-right symmetry breaking}
\label{sec:sugra}
The possibility that left-right symmetry may remain unbroken to
low scales, and such breaking may be compatible with standard 
cosmology has been studied in an earlier work \cite{Mishra:2008be}.
Specifically it was examined whether in the supersymmetric left-right
symmetric model the parity breaking could be of hidden sector origin,
and communicated to visible sector through gauge mediation at a low scale.
The attempt is to see if several of the puzzles of the Standard Model
and incorporation of right handed neutrinos is essentially possible
within a few orders of magnitude of the TeV scale. It was found that
messengers of a particular implementation of gauge mediated supersymmetry 
breaking can also communicate  left-right symmetry violation. This is
independent of the mechanism for the left-right symmetry violation in 
the hidden  sector, the origin of which therefore remains unknown. 

The question is, does supergravity have the potential to address the
origin of left-right symmetry violation at a low scale?
There is a folk theorem that discrete symmetries can be violated
by quantum gravity effects. The reasoning runs as follows.
Formation of black hole horizons can cause unaccounted violation of a 
global charge, while preserving gauge charges. We then expect black hole-like
virtual states in quantum gravity which can induce effective terms violating
the global charges. Such induced terms however do not arise in the
process of perturbative renormalization process, since every perturbative
term, even in a non-renormalisable lagrangian should obey the expected
symmetries. The effective terms would therefore arise from instanton-like
effects. 

Due to discrete nature of the symmetry, the signal of its breaking
would be in the difference in the coefficients of the terms that get
interchanged under the symmetry operation.
The structure of supergravity ensures that at the renormalisable level 
gravity couples 
separately to the left sector and right sector with no mixing terms. 
The field contents are identical and the gauge  couplings in the two 
sectors are identical at this order. It appears very difficult to see 
how supergravity would distinguish between the constants induced in 
the two sectors. We also note that $N=1$ supergravity  is finite at 
one-loop level.  This leads us to suspect that we should not  expect 
parity violating terms from supergravity, at least in the leading  
order in $1/M_{Pl}$. 

Thus a justification for considering $1/M_{Pl}$ terms differing in their
coefficients arises from the possibility that such are a result of
gravity mediated supersymmetry breaking communicated from the hidden sector. 
For this to work we must  assume one of two possibilities, either that  
the gauge group governing  the hidden sector does not admit left-right 
symmetry as a subgroup or that such symmetry is broken in the hidden sector. 
The breaking  should then be communicated to the visible sector along 
with the supersymmetry breaking. The root cause of the parity breaking
then would remain hidden as in our earlier work. We shall also consider 
the next to leading order terms and find that they may suffice only 
marginally to solve the problem of unacceptable domain walls in cosmology.

Regardless of their origin, the structure of the symmetry breaking terms 
in the scalar potential will be similar to that of the terms  that can be 
derived from the superpotential, as happens in the case of soft supersymmetry 
breaking terms \cite{Martin:1997ns}. Thus we may use the usual supergravity 
formalism to derive the terms through which Planck suppressed left-right
symmetry breaking may get manifested.
A similar approach has been adopted in the context of MSSM 
in \cite{Antoniadis:2008ur}  
where the origin and the effect of higher dimensional operators 
has been discussed in the context of collider data.
In the remainder of this section we summarize the essential formalism 
to be used in our calculation. We adopt the notation described in \cite{Martin:1997ns}.
The supergravity lagrangian is obtained from three functions of
complex scalar fields, viz. superpotential $(W)$, K\"{a}hler potential
$(K)$ and gauge kinetic function $f_{ab}$. 
The $F$-term contribution to the scalar potential in supergravity theory
\begin{equation}
 V_F = k_i^j F_j F^{*i} - 3 e^{\frac{K}{M_{Pl}^2}}W W^*/M_{Pl}^2
\label{eqn:scalarpot}
\end{equation}
where
\begin{equation}
F^i = -\left[ (K^{-1})_l^i \left( W^{*l} + W^* K^l/M_{Pl}^2\right) \right] 
\label{eq:F-term}
\end{equation}
Making use of Eq.(\ref{eq:F-term}) in Eq.(\ref{eqn:scalarpot})
the individual terms in $V_F$ can be written as,
\begin{equation}
 V_F = (K^{-1})^{*k}_{\phantom{*k}{l}}
\left[W^*_k W^l + \frac{W^*_k W K^l}{M_{Pl}^2}+\frac{W^*K_kW^l}{M_{Pl}^2}
+\frac{W^* K_k W K^l}{M_{Pl}^4}\right] + {\rm Higher~order~terms}
\label{eqn:V_F}
\end{equation}
Here we have considered the first term of Eq.(\ref{eqn:scalarpot}).
The scalar potential contains D-term contributions from gauge interactions
which are given by
\begin{equation}
V_D=\frac{1}{2}Re[f^{-1}_{ab} \hat{D}_a\hat{D}_b]
\label{eqn:V_D}
\end{equation}
where \begin{equation}
 \hat{D}_a=-K^i(T^a)_i^{\phantom{i}{j}}\phi_j=
-\phi^{*j}(T^a)_j^{\phantom{j}{i}}K_i
\label{eqn:Dterms}
\end{equation}
and $f^{ab}$ is the gauge kinetic function which is given by
\begin{equation}
 f^{ab} = \delta_{ab}[1/g_a^2 +f_a^i\phi_i/M_{Pl}+....]
\end{equation}
For our purpose it will be sufficient to consider $f^{ab}=\delta_{ab}/g_a^2$
since we do not expect left-right asymmetry to arise from the gauge sector.
In the following we shall consider the terms arising in the scalar potential
from expanding  $W$ and $K$ in the powers of $(1/M_{Pl})$.

\section{Gravity induced operators in Left-Right Supersymmetric Model}
\label{sec:lrsm}

The key difference in any realistic model in comparison with the 
generic considerations of \cite{Rai:1992xw} is that gauge invariance
restricts the structure of the non-renormalizable terms.  For instance 
in \cite{Lew:1993yt} the operators considered were  
\begin{equation}
 V_{\textrm{non-SUSY}} \sim c_1 \frac{1}{M_{Pl}^2}(\Delta_L^\dag \Delta_L)^3
+ c_2 \frac{1}{M_{Pl}^2}(\Delta_R^\dag \Delta_R)^3
\label{eq:nonsusyparitybr}
\end{equation}
In other words, gauge invariance requires the terms to be $O(1/M_{Pl}^2)$ 
rather than $O(1/M_{Pl})$.
A difference in pressure caused by such operators, after putting vacuum 
expectation values of the fields $\Delta \sim M_R$, would be adequate
to remove the domain walls accompanied by radiation dominated era evolution,
eq. (\ref{eq:eps-vsix}), and even more so the domain walls that evolved 
during an effective matter dominated era, eq. (\ref{eq:eps-v11half}).

But in the models we consider, supersymmetry forbids the kind of terms shown
in eq. (\ref{eq:nonsusyparitybr}); on the other hand, generic parity breaking 
terms of $O(1/M_{Pl})$  respecting $SU(2)_L$ and $SU(2)_R$ gauge invariance are 
permitted. This is due to additional field content with different gauge
charges. This gain in order of $1/M_{Pl}$ is however offset by the fact that
due to demands of phenomenology, some of the fields can 
acquire TeV scale vacuum expectation values as well. 
Here we have studied two successful supersymmetric implementation of left-right 
symmetry but the method can be extended to other implementations. 

The minimal left-right SUSY model is based on the gauge group
$SU(3)_c$ $\otimes ~SU(2)_L$ $\otimes ~SU(2)_R$ $\otimes ~U(1)_{B-L}$.
The anomaly free $B-L$ global symmetry of the Standard Model is promoted to 
a gauge symmetry. The quark, lepton and Higgs fields for the minimal left-right SUSY
model, with their respective quantum numbers under the gauge group
$SU(3)_c$ $\otimes ~SU(2)_L$ $\otimes ~SU(2)_R$ $\otimes ~U(1)_{B-L}$ are
given by,
\begin{eqnarray}
Q = (3,2,1,1/3), & \quad & Q_c = (3^*,1,2,-1/3), \nonumber  \\
L = (1,2,1,-1),  & \quad & L_c = (1,1,2,1), \nonumber \\
\Phi_i = (1,2,2,0),    & \quad & \textrm{for } i = 1,2, \nonumber \\
\Delta = (1,3,1,2),    & \quad & \bar{\Delta} = (1,3,1,-2), \nonumber \\
\Delta_c = (1,1,3,-2), & \quad & \bar{\Delta}_c = (1,1,3,2). 
\label{eq:minimalset}
\end{eqnarray}
where we have suppressed the generation index for simplicity of notation.
In the Higgs sector, the bidoublet $\Phi$ is doubled to have nonvanishing 
Cabbibo-Kobayashi-Maskawa matrix, whereas the $\Delta$ triplets are doubled
to have anomaly cancellation. Under discrete parity symmetry the fields
are prescribed to transform as,
\begin{eqnarray}
Q \leftrightarrow Q_c^*, \quad & 
L \leftrightarrow L_c^*, \quad & 
\Phi_i \leftrightarrow \Phi_i^\dagger,  \nonumber \\
\Delta \leftrightarrow \Delta_c^*,  \quad & 
\bar{\Delta} \leftrightarrow \bar{\Delta}_c^*.
\label{eq:parity} 
\end{eqnarray}
However, this minimal left-right symmetric model is unable
to break parity spontaneously \cite{Kuchimanchi:1993jg, Kuchimanchi:1995vk}.
Inclusion of nonrenormalizable terms gives a more realistic structure of
possible vacua \cite{Mohapatra:1995xd,Aulakh:1998nn,Aulakh:1997fq}. 
Such terms were studied for the case when the scale 
of $SU(2)_R$ breaking is high, close to Planck scale. 

\subsection{The ABMRS model with a pair of triplets}
\label{subsec:ABMRSmodel}
Due to the difficulties with the model discussed above, a ``minimal''
renormalisable model was developed early in \cite{Aulakh:1997ba,
Aulakh:1998nn,Aulakh:1997fq}, and will be referred to here as the ABMRS model. 
In this  model two triplet 
fields $\Omega$ and $\Omega_c$, were added, with the following quantum numbers,
\begin{equation}
\Omega = (1,3,1,0), \quad \Omega_c = (1,1,3,0),
\end{equation} 
which was shown to improve the situation with only the renormalisable
terms \cite{Aulakh:1997ba, Aulakh:1997fq, Aulakh:2003kg}.
It was shown that this model breaks down to minimal 
supersymmetric standard model (MSSM) at low scale.
This model was studied in the context of cosmology in 
 \cite{Yajnik:2006kc, Sarkar:2007ic} specifically, the mechanism for 
leptogenesis via Domain Walls in \cite{Sarkar:2007er}.

The superpotential for this model including Higgs fields only
is given by,
 
\begin{eqnarray}
W_{LR} &=& m_\Delta  {\rm  Tr}\, \Delta \bar{\Delta} 
  + m_\Delta  {\rm Tr}\,\Delta_c \bar{\Delta}_c
  + \frac{m_\Omega}{2} {\rm Tr}\,\Omega^2 
  + \frac{m_\Omega}{2} {\rm Tr}\,\Omega_c^2 \nonumber \\
&& + ~\mu_{ij} {\rm Tr}\,  \tau_2 \Phi^T_i \tau_2 \Phi_j 
  + a {\rm Tr}\,\Delta \Omega \bar{\Delta}
  + a {\rm Tr}\,\Delta_c \Omega_c \bar{\Delta}_c \nonumber \\
& &  + ~\alpha_{ij} {\rm Tr}\, \Omega  \Phi_i \tau_2 \Phi_j^T \tau_2 
  + \alpha_{ij} {\rm Tr}\, \Omega_c  \Phi^T_i \tau_2 \Phi_j \tau_2 ~.
\label{eq:AMSsuperpot}
\end{eqnarray}
Since supersymmetry is broken
at a very low scale, we can employ the $F$ and $D$ flatness conditions
obtained from the superpotential, to get a possible solution for the 
vacuum expectation values (vev's)
for the Higgs fields. 
\begin{equation} 
\begin{array}{ccc}
\langle \Omega \rangle = 0, & \qquad
\langle \Delta \rangle = 0, & \qquad
\langle \bar{\Delta} \rangle = 0, \\ [0.25cm]
\langle \Omega_c \rangle = 
\begin{pmatrix}
\omega_c & 0 \\
0 & - \omega_c
\end{pmatrix}, & \qquad
\langle \Delta_c \rangle =
\begin{pmatrix}
0 & 0 \\
d_c & 0
\end{pmatrix}, & \qquad
\langle \bar{\Delta}_c \rangle =
\begin{pmatrix}
0 & \bar{d}_c \\
0 & 0
\end{pmatrix}.
\label{eq:rhvev} 
\end{array}
\end{equation}
This solution set is of course not unique. Since the original theory is
parity invariant a second  solution for the $F$ and $D$ flat
conditions exists, with Left type fields' vev's exchanged with those 
of the Right type fields \cite{Sarkar:2007ic, Sarkar:2007er}. 

With vev's as in eq. (\ref{eq:rhvev}) the pattern of breaking is 
\begin{eqnarray}
SU(2)_L \otimes SU(2)_R \otimes U(1)_{B-L} &\stackrel{M_R}{\longrightarrow}& 
SU(2)_L \otimes U(1)_R \otimes U(1)_{B-L}\\
&\stackrel{M_{B-L}}{\longrightarrow}& SU(2)_L \otimes U(1)_Y
\end{eqnarray}
The model introduces a new mass scale, $m_\Omega$. However, 
it was observed \cite{Aulakh:1997fq} that these terms can be forbidden in the
superpotential by invoking an $R$ symmetry, and then the corresponding terms 
appearing in the scalar potential can be  interpreted as soft terms entering only 
after supersymmetry breakdown at the electroweak scale. This approach
imposes a condition on the scales of breaking, with respect to the electroweak 
scale $M_W$,
\begin{equation}
M_R M_W \simeq M_{B-L}^2
\end{equation}
This relation raises the interesting possibility that the scale
of $M_R$ can be as low as $10^4$ to $10^6$ GeV, with corresponding
very low scale $10^3$ to $10^4$GeV of lepton number violation, 
opening the possibility of low energy leptogenesis
\cite{Sahu:2004sb,Sarkar:2007er}.

As discussed in Sec.(\ref{sec:sugra}) we shall proceed to find the
$1/M_{Pl}$ terms in the effective potential by expanding K\"{a}hler
potential and superpotential in powers of $1/M_{Pl}$. Here we include
$\Delta(\bar{\Delta}),\Delta_c(\bar{\Delta}_c)$ and $\Omega(\Omega_c)$
fields in the expansion of the K\"{a}hler potential and superpotential.
The K\"{a}hler potential in this model upto $(1 / M_{Pl})$ can be written as,
\begin{eqnarray} 
K &=& {\rm Tr}\,(\Delta \Delta^{\dagger} +
\bar{\Delta} \bar{\Delta}^{\dagger}) +
{\rm Tr}\,( \Delta_c \Delta_c^{\dagger}+
\bar{\Delta}_c \bar{\Delta}_c^{\dagger})
+ {\rm Tr}\, (\Omega \Omega^\dagger )+{\rm Tr}\ (\Omega_c\Omega_c^\dagger)\nonumber\\
&+& \frac{C_L}{M_{Pl}} ({\rm Tr}\,\Delta \Omega \Delta^{\dagger}
+ {\rm Tr}\,\bar{\Delta} \Omega \bar{\Delta}^{\dagger} + h.c)+
\frac{C_R}{M_{Pl}} ({\rm Tr}\, \Delta_c \Omega_c \Delta_c^{\dagger}
+ {\rm Tr}\, \bar{\Delta}_c \Omega_c \bar{\Delta}_c^{\dagger} + h.c).
\label{eqn:AMSkahler} 
 \end{eqnarray}
Where $C_L$ and $C_R$ are dimensionless constants.
The superpotential can also be expanded in powers of $(1 / M_{Pl})$ 
which is written as 
\begin{eqnarray}
 W &=& W_{ren}  + W_{nr}  \nonumber \\
&=& m_\Delta  ({\rm  Tr}\, \Delta \bar{\Delta} 
  + {\rm  Tr}\,\Delta_c \bar{\Delta}_c)
  + \frac{m_\Omega}{2} ({\rm Tr}\,\Omega^2 
  + {\rm  Tr}\,\Omega_c^2) + a ({\rm Tr}\,\Delta \Omega \bar{\Delta}
  + {\rm  Tr}\,\Delta_c \Omega_c \bar{\Delta}_c) \nonumber \\
&+& \frac{a_L}{2M_{Pl}} ({\rm  Tr}\Delta \bar{\Delta})^2
 + \frac{a_R}{2M_{Pl}} ({\rm  Tr}\Delta_c \bar{\Delta}_c)^2
 + \frac{b_L}{M_{Pl}} {\rm  Tr}\Delta^2 {\rm  Tr}\,\bar{\Delta}^2
+  \frac{b_R}{M_{Pl}} {\rm  Tr}\Delta_c^2 {\rm  Tr}\,\bar{\Delta}_c^2 \nonumber \\
&+& \frac{c_L}{4M_{Pl}} ({\rm  Tr}\Omega^2)^2 + \frac{c_R}{4M_{Pl}}
({\rm  Tr} \Omega_c^2)^2 + \frac{d_L}{2M_{Pl}}{\rm  Tr}\Omega^2{\rm  Tr}\Delta \bar{\Delta}
+ \frac{d_R}{2M_{Pl}}{\rm  Tr}\Omega_c^2{\rm  Tr}\Delta_c \bar{\Delta}_c\nonumber\\
&+& \frac{f}{M_{Pl}}{\rm  Tr}\Delta\bar{\Delta}{\rm  Tr}\Delta_c\bar{\Delta}_c
+\frac{h_1}{M_{Pl}}{\rm  Tr}\Delta^2 {\rm  Tr}\Delta_c^2+\frac{h_2}{M_{Pl}}
{\rm  Tr}\,\bar{\Delta}^2{\rm  Tr}\,\bar{\Delta}_c^2\nonumber \\
 &+& \frac{j}{M_{Pl}}{\rm  Tr}\Omega^2{\rm  Tr} \Omega_c^2
+\frac{k}{M_{Pl}}{\rm  Tr}\Omega^2{\rm  Tr}\Delta_c \bar{\Delta}_c
+\frac{m}{M_{Pl}}{\rm  Tr} \Omega_c^2{\rm  Tr}\Delta \bar{\Delta}
\label{eq:ams-nr-W}
\end{eqnarray}
where the coefficients appearing against the individual terms are
dimensionless constants.

The effective potential resulting from above modifications is calculated 
in appendix \ref{sec:ap1}.
We have also examined the D-terms in the effective potential and find that
they do not give rise to $O(\frac{1}{M_{Pl}})$ terms. Thus we only
need to deal with $F$-terms. Assuming a phase in which the right-type fields 
get a non-trivial vev and all left-type  fields 
have zero vev, the expression for the leading term in $1/M_{Pl}$ the scalar 
potential becomes
\begin{equation}
 V^R_{eff} \sim \frac{a(c_R + d_R)}{M_{Pl}}M_R^4M_W + \frac{a(a_R + d_R)}{M_{Pl}}
M_R^3M_W^2
\end{equation}
Now due to left-right symmetry, there is also a corresponding phase
in which the left-type fields getting a vev, but no the right-type fields. 
For this phase the value of the effective potential is 
\begin{equation}
 V^L_{eff} \sim \frac{a(c_L + d_L)}{M_{Pl}}M_R^4M_W + \frac{a(a_L + d_L)}{M_{Pl}}
M_R^3M_W^2
\end{equation}
The possibility of these two phases of approximately equal energy density 
gives rise to domain walls separating such phases.
The pressure difference across the walls is proportional to the 
difference in energy density between two sides of the wall, and is 
given by,
\begin{eqnarray}
 \delta\rho&\sim&[(c_L-c_R)+(d_L-d_R)] \frac{M_R^4M_W}{M_{Pl}}
 + [(a_L-a_R)+(d_L-d_R)] \frac{M_R^3M_W^2}{M_{Pl}}\nonumber\\
&\sim& \kappa^A \frac{M_R^4M_W}{M_{Pl}}+{\kappa'}^A\frac{M_R^3M_W^2}{M_{Pl}}
\label{eq:ams-kapa}
\end{eqnarray}
Where $\kappa^A=(c_L-c_R)+(d_L-d_R)$ and $\kappa'^A=(a_L-a_R)+(d_L-d_R)$,
and the superscript $A$ refers to the ABMRS model. 
From Eq.(\ref{eq:ams-kapa}), we see that to leading order in $1/M_{Pl}$
there are two kinds of operators appearing in $\delta\rho$, differing 
in powers of $(M_W/M_R)$. 

We shall now compare these operators
with the energy density required for the successful removal of the domain
walls in the three cases to be labelled as RD, MD and WI respectively, 
discussed in Sec.(\ref{sec:dicsymmcosmo}).
Comparing  Eq.(\ref{eq:eps-vsix}), with individual operators in 
Eq.(\ref{eq:ams-kapa}) and taking the scale $M_R$ as $10^6$GeV,
and taking the more dominant term $\kappa$, we get the constraint 
\begin{equation}
 \kappa^A_{RD} > 10^{-10} \left( \frac{M_R}{10^6 {\rm GeV}}\right)^2, \
\end{equation}
This is easily satisfied at low scale of $M_R$ proposed.  For $M_R$ 
scale tuned to $~10^9$GeV  needed to avoid gravitino problem after 
reheating at the end of inflation, $\kappa_{RD}\sim 10^{-4}$, a 
reasonable constraint. but requires
$\kappa^A_{RD}$ to be $O(1)$ if the scale of $M_R$ is an
intermediate scale $10^{11}$GeV.

Next, comparing Eq.(\ref{eq:eps-v11half}) 
with individual terms in Eq.(\ref{eq:ams-kapa}), the constraint on 
$\kappa^A_{MD}$ 
is found to be
\begin{equation}
 \kappa^A_{MD} > 10^{-2} \left( \frac{M_R}{10^6 {\rm GeV}}\right)^{3/2},
\label{eq:kappaAMD}
\end{equation}
which puts a modest requirement on the value of $\kappa_{RD}$ for
suitable disappearance of domain walls. However taking $M_R\sim10^9$GeV 
being the temperature scale required to have thermal leptogenesis
without the undesirable gravitino  production, leads to $\kappa_{MD}>10^{5/2}$,
an unacceptable requirement.
The MD case is in fact generic to supersymmetric and string inspired models,
\cite{Coughlan:1983ci,Banks:1993en,deCarlos:1993jw}
due to moduli production. And we find that in this case the ABMRS model 
requires a low scale of $M_R$ and non-thermal or resonant leptogenesis. 

In the WI case, eq.(\ref{eq:deltaRhoTD}) there is extreme 
sensitivity to the scales of $M_R$ and $T_D$. Proceeding in same way as 
above comparing Eq.(\ref{eq:deltaRhoTD}) with Eq.(\ref{eq:ams-kapa}) 
the constraint on $\kappa$ 
is found to be,
\begin{equation}
 \kappa^A_{WI} > 10^{-4} \left( \frac{10^6 {\rm GeV}}{M_R}\right)^{15}
\left( \frac{T_D}{10{\rm GeV} }\right)^{12},
\end{equation}
This is a reasonable constraint for the proposed median values of the
two mass scales. However the constraint makes the model rather strongly 
predictive. The scale of decay of the wall complex $T_D$ can be any
value below chosen $M_R$ scale. Thus if $T_D\sim10^4$GeV, then 
$M_R$ is forced to be closer to the gravitino scale $10^9$Gev.
This can be problematic if the reheating temperature after the
disappearance of the domain walls in comparable to the temperature
scale of the original phase transition. The Universe would reheat to
$10^9$Gev, raising the possibility of unacceptable gravitino regeneration.

Finally we consider the possibility raised in sec. \ref{sec:sugra}, that
the gravity induced terms are of direct origin, and due to one-loop
finiteness of supergravity, do not give rise to $1/M_{Pl}$ terms in
the superpotential or the K\"ahler potential. In such a case
the most dominant operator to be considered is suppressed by $(1/M_{Pl})^2$.
In the ABMRS case we find, after substituting the vacuum expectation 
values, that such an operator has the magnitude 
\begin{equation}
\delta V_{\textrm{Next-to-leading-order}} \sim \frac{M_R^4 M_W^2}{M_{Pl}^2}
\label{eq:nexttoleading}
\end{equation}
So long as we are considering theories with $M_R$ values less that
an intermediate scale $10^{11}$GeV, such terms are subdominant to the
ones considered above, tightening each of the above constraints by a
factor $M_{Pl}/M_W \sim 10^{16}$. Such a constraint immediately renders
the first two scenarios of domain wall evolution cosmologically unacceptable.
The third case of weak inflation however continues to be possible for
phenomenologically acceptable values of the energy scales.

To summarise the situation for the ABMRS model, we have found that
there is an upper bound on the scale $M_R$ if the wall evolution unfolds
during a radiation dominated epoch or a matter dominated epoch. In the 
the latter case, which is generic for string theory cosmology with presence 
of heavy moduli fields, the natural value of $M_R$ is required to be 
significantly lower than, $10^9$GeV. In the case of an evolution
accompanied by weak inflationary epoch, there is no upper bound, rather
a lower bound on the scale $M_R$ but which is extremely sensitive to
the value of the scale $T_D$ at which the walls may finally disappear.

\subsection{The BM model with a single singlet}
\label{subsec:BMmodel}
An independent approach to improve the minimal model with
introduction of a parity odd singlet \cite{Cvetic:1985zp}, was adopted
in \cite{Kuchimanchi:1993jg, Kuchimanchi:1995vk}.
However this was shown at tree level to lead to charge-breaking vacua 
being at a lower potential than charge-preserving vacua.

Recently, an alternative to this has been considered in \cite{Babu:2008ep} 
here a  superfield $S = (1,1,1,0)$ also singlet under parity is included
in addition to the minimal set of Higgs required as in
Eq.(\ref{eq:minimalset}). The $\Delta_c,\bar{\Delta}_c$ are required
for $SU(2)_R$ $\otimes ~U(1)_{B-L}$ symmetry breaking without
inducing R-parity violating couplings. The singlet field $S$ is 
introduced so that $SU(2)_R$ $\otimes ~U(1)_{B-L}$ symmetry breaking
occurs in the supersymmetric limit. The superpotential
including the Higgs fields is  given by,

The superpotential is  given by,
\begin{center} 
$ W_{LR} = W^{(1)} + W^{(2)}$
 \end{center} 
Where
\begin{eqnarray} 
 W^{(1)} &=& {\bf h}_l^{(i)} L^T \tau_2 \Phi_i \tau_2 L_c 
 + {\bf h}_q^{(i)} Q^T \tau_2 \Phi_i \tau_2 Q_c\nonumber\\ 
&& + ~ i {\bf f^\ast} L^T \tau_2 \Delta L 
 + ~i {\bf f} L^{cT}\tau_2 \Delta_c L_c \nonumber \\
&& + ~ S~[~\lambda^\ast ~ {\rm Tr}\,\Delta  \bar{\Delta}
  + ~ \lambda ~ {\rm Tr}\,\Delta_c  \bar{\Delta}_c
  + ~ \lambda^\prime_{ab} {\rm Tr}\, \Phi_a^T \tau_2 \Phi_b \tau_2 - M_R^2~]
\label{eq:Wone}\\
W^{(2)} &=& ~M_\Delta  {\rm  Tr}\, \Delta \bar{\Delta} 
  + M_\Delta^\ast  {\rm Tr}\,\Delta_c \bar{\Delta}_c\nonumber\\
&& + ~\mu_{ab} {\rm Tr}\, \Phi^T_a \tau_2 \Phi_b \tau_2
   + M_s S^2 + \lambda_s S^3
\label{eq:Wtwo}
 \end{eqnarray} 

In this analysis the terms in $W^{(2)}$ has been assumed to be zero.
Dropping the terms in $W^{(2)}$ makes the theory more symmetric 
and more predictive. 
It is observed that dropping quadratic and cubic terms in
$S$ leads to an enhanced $R$-symmetry.
Further, dropping the massive couplings introduced for $\Delta$'s 
means that $\Delta$ masses arise purely from SUSY breaking effects, 
keeping these fields light and relevant to collider phenomenology.
Dropping the $\mu_{ab}$ terms for $\Phi$ fields makes it possible to 
explain the $\mu$ parameter of MSSM as being spontaneously induced 
from $S$ vev through terms in $W^{(1)}$.
Additionally, absence of the $W^{(2)}$ terms can be shown to 
solve the SUSY CP and strong CP problems.

The presence of linear terms in 
$S$ in $W^{(1)}$ makes possible the following SUSY vacuum,
\begin{equation}
\vev{S} = 0, \quad \lambda v_R \bar{v}_R + \lambda^\ast
v_L \bar{v}_L = M_R^2
\label{eq:SSvev} 
\end{equation}
where $v_L(\bar{v}_L)$ and $v_R(\bar{v}_R)$ are the vev's of the
neutral components of $\Delta(\bar{\Delta})$  and $\Delta_c
(\bar{\Delta}_c)$ fields respectively. From Eq.(\ref{eq:SSvev})
it is clear that we have a flat direction in  the $v_L$ - $v_R$ space.
Assuming that the flat directions are lifted, we have two choices viz.
\begin{equation}
v_R=\bar{v}_R=0, \qquad |v_L|=|\bar{v}_L|=\frac{M_R}{\sqrt{\lambda^\ast}}
\label{eq:SSRsol}
\end{equation}

\begin{equation}
v_L=\bar{v}_L=0, \qquad |v_R|=|\bar{v}_R|=\frac{M_R}{\sqrt{\lambda}}
\label{eq:SSLsol}
\end{equation}

The important result is that after SUSY breaking and emergence of 
SUSY breaking soft terms,  integrating out heavy sleptons modifies
the vacuum structure due to Coleman-Weinberg type one-loop terms 
which must be treated  to be of same order as the other terms in 
$V^{eff}$. Accordingly, it is shown\cite{Babu:2008ep} that the 
$V^{eff}$ contains terms of the form
\begin{equation} 
 V_{1-loop}^{eff} (\Delta_c) \sim - |f|^2 m_{L^c}^2
{\rm  Tr}\, (\Delta_c \Delta_c^\dagger) A_1^R - |f|^2 m_{L^c}^2
{\rm  Tr}\, (\Delta_c \Delta_c)
{\rm  Tr}\, (\Delta_c^\dagger \Delta_c^\dagger) A_2^R
\label{eq:onelR}
 \end{equation} 
where $A_1^R$ and $A_2^R$ are constants obtained from expansion of the
effective potential. Presence of these terms is shown to lead to
the consequence of a preference  for electric charge-preserving
vacuum over the charge-breaking  vacuum, provided $m_{L^c}^2 < 0$.

Further Eq.(\ref{eq:SSRsol}) also constitutes a valid solution
of Eq.(\ref{eq:SSvev}). In this vacuum the soft terms can
give rise to the following terms in the effective potential,
\begin{equation} 
 V_{1-loop}^{eff} (\Delta) \sim - |f|^2 m_L^2
{\rm  Tr}\, (\Delta \Delta^\dagger) A_1^L - |f|^2 m_L^2
{\rm  Tr}\, (\Delta \Delta) {\rm  Tr}\, (\Delta^\dagger\Delta^\dagger)A_2^L
\label{eq:onelL} 
\end{equation} 
with $A_1^L$ and $A_2^L$ constants.
Thus the choice of known phenomenology is only one of two possible local
choices, and formation of domain walls is inevitable.

Here we analyse the full superpotential without setting $W^{(2)}$ 
terms zero. Like in the ABMRS model, we study here the scalar potential 
taking supergravity into account. Including triplet fields and singlet field
$S$, the K\"{a}hler potential in this model
upto $O(1 / M_{Pl})$ is given by,
\begin{eqnarray} 
K &=&  {\rm Tr}\,( \Delta \Delta^{\dagger} +
\bar{\Delta}  \bar{\Delta}^{\dagger})
+ {\rm Tr}\, (\Delta_c \Delta_c^{\dagger}+
\bar{\Delta}_c \bar{\Delta}_c^{\dagger}) + |S|^2
+ \frac{C_L}{M_{Pl}}( {\rm Tr}\, \Delta S \Delta^{\dagger} 
+ {\rm Tr}\,\bar{\Delta}  S \bar{\Delta}^{\dagger}+  h.c) \nonumber\\
&& + \frac{C_R}{M_{Pl}}( {\rm Tr}\, \Delta_c S \Delta_c^{\dagger}
+ {\rm Tr}\,\bar{\Delta}_c S \bar{\Delta}_c^{\dagger} +  h.c)
+\frac{d}{M_{Pl}}(S^3+h.c.).
\label{eqn:BMkahler} 
\end{eqnarray}
The superpotential upto $(1/M_{Pl})$ order is given by,
\begin{eqnarray}
 W &=& W_{ren}  + W_{nr}  \nonumber \\
&=& m_\Delta  ({\rm  Tr}\, \Delta \bar{\Delta} 
  + {\rm  Tr}\, \Delta_c \bar{\Delta}_c)
   + M_s S^2 + \lambda_s S^3 +
S[\lambda^* {\rm  Tr}\, \Delta \bar{\Delta}+
\lambda{\rm  Tr}\Delta_c \bar{\Delta}_c-M_R^2]\nonumber \\
&+& \frac{a_L}{2M_{Pl}} ({\rm  Tr}\Delta \bar{\Delta})^2
 + \frac{a_R}{2M_{Pl}} ({\rm  Tr}\Delta_c \bar{\Delta}_c)^2
 + \frac{b_L}{2M_{Pl}} {\rm  Tr}\Delta^2 {\rm  Tr}\,\bar{\Delta}^2
+  \frac{b_R}{2M_{Pl}} {\rm  Tr}\Delta_c^2 {\rm  Tr}\,\bar{\Delta}_c^2\nonumber\\
 &+&\frac{c}{M_{Pl}}S^4 + \frac{c_L}{M_{Pl}} S^2 {\rm  Tr}\, \Delta \bar{\Delta}
+ \frac{c_R}{M_{Pl}} S^2 {\rm  Tr}\Delta_c \bar{\Delta}_c\nonumber\\
&+& \frac{f}{M_{Pl}}{\rm  Tr}\, \Delta \bar{\Delta} {\rm  Tr}\Delta_c \bar{\Delta}_c
+ \frac{h_1}{M_{Pl}}{\rm  Tr}\Delta^2 {\rm  Tr}\Delta_c^2+
\frac{h_2}{M_{Pl}}{\rm  Tr}\,\bar{\Delta}^2{\rm  Tr}\,\bar{\Delta}_c^2
\label{eq:bm-nr-W}
\end{eqnarray}
The effective potential has been calculated in the appendix(\ref{sec:ap2})
considering the term from the effective potential $(K^{-1})^{*k}_{\phantom{*k}{l}}
W^*_k W^l$.  When the right-type fields get a vev the scalar potential can be
written as,
\begin{equation}
 V^R_{eff}\sim \frac{a_R}{M_{Pl}}M_R^5 + \frac{a_R}{M_{Pl}} s M_R^4
+ \frac{C_R}{M_{Pl}} s^2 M_R^3 +  \frac{C_R}{M_{Pl}} s^3 M_R^2
\end{equation}
where $s$ is the scale at which $S$ gets a vev.
To calculate the potential when only the left type fields get 
vacuum expectation values, we introduce corresponding coefficients $a_L$, 
etc. We then compute the pressure difference across the walls as
\begin{eqnarray}
 \delta\rho&\sim& (a_L-a_R) \frac{M_R^5}{M_{Pl}} + \ldots
+ (C_L-C_R) \frac{s^3 M_R^2}{M_{Pl}}\nonumber\\
&\sim& \kappa^B \frac{M_R^5}{M_{Pl}} + \ldots
+{\kappa'}^B\frac{s^3 M_R^2}{M_{Pl}}
\label{eq:bm-kapa}
\end{eqnarray}
where $\kappa^B=(a_L-a_R)$, ${\kappa'}^B=(C_L-C_R)$, superscript $B$ referring to
the BM model and $\ldots$ is in lieu of terms which, as we explain next, are 
relatively unimportant.
According to BM model, the value $s$ is of the scale of supersymmetry breaking.
If this scale is TeV scale, then $M_R$ is expected to be higher and the
$\kappa'^B$ terms are subdominant. In case however the supersymmetry breaking scale
is $10^{11}$ GeV, it could be higher then the scale of $M_R$. In this case 
the $\kappa'^B$ term is expected to dominate.

In the case of TeV scale supersymmetry breaking, comparing the first
term in Eq.(\ref{eq:bm-kapa})  with Eqs.(\ref{eq:eps-vsix}),
(\ref{eq:eps-v11half}) and (\ref{eq:deltaRhoTD}) of with sec.(\ref{sec:dicsymmcosmo}), 
we get the corresponding constraints on  possible values of $\kappa$ as
\begin{eqnarray}
 \kappa^B_{RD} &>& 10^{-13} \left( \frac{M_R}{10^6 {\rm GeV}}\right),\\
\kappa^B_{MD} &>& 10^{-6} \left( \frac{M_R}{10^6 {\rm GeV}}\right)^{1/2},\\
 \kappa^B_{WI} &>& 10^{-8} \left( \frac{10^6 {\rm GeV}}{M_R}\right)^{16}
\left( \frac{T_D}{10{\rm GeV} }\right)^{12}
\end{eqnarray}
Thus for the proposed $M_R$ scale of $10^6$GeV there is no serious constraint
on $\kappa^B$ values. Only in the scenario with weak inflation, if $T_D$ scale
is high, such as $100$GeV, the value $M_R\sim 10^6$GeV becomes marginal,
but due to large powers of the mass scale present, a small increase
in $M_R$ easily offsets the effect of the increase in $T_D$. Overall, the
kind of operators obtained in this particular model provide no constraint
on the mass scale $M_R$, as long as the scale of supersymmetry breaking
is TeV scale.

To check other possibilities, we consider the supersymmetry
breaking scale $M_S$ and hence $s$ to be $\sim 10^{11}$GeV. In this case we
get, proceeding as above, 
\begin{eqnarray}
\kappa_{RD}'^B &>&  10^{-25} \left( \frac{M_R}{10^6 {\rm GeV}}\right)^4 
\left( \frac{10^{11} {\rm GeV}}{s}\right)^3 \\ 
\kappa_{MD}'^B &>&  10^{-19} \left( \frac{M_R}{10^6 {\rm GeV}}\right)^{7/2}
\left( \frac{10^{11} {\rm GeV}}{s}\right)^3 \\
 \kappa_{WI}'^B &>&  10^{-44} \left(\frac{T_D}{10 {\rm GeV}}\right)^{12}
\left( \frac {10^6 {\rm GeV}}{M_R}\right)^{17}
\left( \frac{10^{11} {\rm GeV}}{s}\right)^3
\end{eqnarray}
This shows that there is no particular constraint on the induced parity 
breaking coefficients due to increase in the scale of supersymmetry breaking
$M_S$, so long as $M_R < s \equiv M_S \sim 10^{11}$GeV.

In summary the BM model remains mostly unrestricted by the present considerations.
This is due to newer terms possible with a gauge singlet.

\section{Conclusion}
\label{sec:concl}
Ever since the discovery of massive neutrinos it has become a tantalising
possibility that the small neutrino masses arise from rich physics at
a high energy scale, which in turn would naturally incorporate right
handed neutrinos. A model that treats this new content symmetrically with
the known contents would naturally lead to the requirement of parity
symmetry in the high energy model. If this discrete symmetry is spontaneously
broken it would lead to formation of domain walls in the early Universe.

We have considered three scenarios for the evolution of transient 
domain wall networks ending in their decay. 
We characterize each model by a dimensionless parameter $\fcal$ which 
is the ratio of the available pressure difference across a wall 
at the time of its decay to the the characteristic energy density $M_R^4$
in the Universe at the time of formation of the wall complex,
see Eqs. (\ref{eq:eps-vsix}), (\ref{eq:eps-v11half}), (\ref{eq:deltaRhoTD}).
Of the three scenarios, the first one unfolds entirely
in a  radiation dominated universe, in which the dynamics is governed 
by the interplay of  forces due to friction and tension and the pressure 
difference across the walls. 
Here we find that the parameter $\fcal$ is given by $(M_R/M_{Pl})^2$. 
The second scenario unfolds in a matter dominated era, where the domain 
wall complex decays as soon as the  energy density of the same dominates 
the energy density of the Universe. The parameter $\fcal$ in this case 
is given by $(M_R/M_{Pl})^{3/2}$. The third 
scenario is an extension of the second one where we assume
that the domain wall network comes to dominate the energy density of
the Universe and continues to do so for a finite epoch before
it decays. Characterizing the required pressure difference
$\delta \rho$ across the domain walls in this case requires additional
input, the ratio of scale factor values $a_{eq}/a_{d}$ 
where subscript $eq$ refers to the epoch at which the domain walls become 
equally as important as the rest of the matter and subscript $d$
refers to the epoch at which the decay of the domain walls occurs.
We characterize this ratio by an equivalent 
``decay temperature'' $T_D$ defined in Eq. (\ref{eq:rhoeq}). 
The formula for $\fcal$ in this case shows very high sensitivity to 
the mass scales concerned.

For each of these cases we study the viability of the two specific models
of spontaneous parity breaking, the ABMRS model, sec. \ref{subsec:ABMRSmodel} 
and the BM model, \ref{subsec:BMmodel}. Each of the particle physics models permits 
intrinsic operators that must match up to the required parameter $\fcal$, 
resulting in final destablisation of the wall complex. If the operators
available in the model cannot provide a $\delta \rho$ of required magnitude, 
the wall complex would not be destabilized,  leading to unacceptable cosmology.

The ABMRS model turns out to be more restrictive, using as it does only non-trivial
representations of the gauge group. In this case high scale of parity
breaking becomes conditionally disfavored, though still viable if
the wall evolution leads to a weak inflationary epoch. The BM model containing
a singlet turns to not be restricted by the considerations here.

Our main conclusion is that a low scale scenario with $M_R\sim10^6$GeV or 
lower is viable and generic. Specifically, in the ABMRS model with domain wall 
evolution in a matter dominated epoch $M_R$ is restricted to remain less than $10^8$GeV,
eq. (\ref{eq:kappaAMD}).
A matter dominated epoch is generic to string theory inspired models with
occurance of moduli fields of mass scale $10^9$GeV 
and hence this restriction is of special interest.

As a broader conclusion we learn that gravity induced higher dimensional operators
can effective in ensuring the removal domain walls generic to a theory with spontaneous 
breakdown of parity, but the result can be model dependent. This mechanism
is then an alternative to an earlier studied \cite{Mishra:2008be} possibility 
of parity breaking mediated by 
the  messengers  in a version of gauge mediated supersymmetry breaking 
scenario. The constraints on the parameters for that scenario
to ensure disappearance of domain walls were rather stringent.
We now see that supergravity induced terms can ensure wall
disappearance with a very modest constraints, which can be predictive
in some cases. The question of what underlying physics results in gravity
inducing parity breaking terms remains open. It appears also tied to whether
the terms arise directly from supergravity in the visible sector or
through the gravity mediated supersymmetry breaking effects.

\section{acknowledgment}

\appendix
\section{AMS Model}
\label{sec:ap1}
The scalar potential contains D-term contributions from gauge interactions.
From Eq.(\ref{eqn:V_D})
\begin{eqnarray}
V_D &=& \frac{1}{2}Re(g_a^2/\delta_{ab}\hat{D}_a\hat{D}_b)\nonumber\\
&=& \frac{g_a^2}{2}Re(\hat{D}_a\hat{D}_a)\nonumber\\
&=& \frac{g_a^2}{2} Re\arrowvert Tr(K_{\Delta}(T^a)\Delta)
+ Tr(K_{\bar{\Delta}}(T^a)\bar{\Delta})+
Tr(K_{\Omega}(T^a)\Omega)\arrowvert^2\nonumber\\
&=& \frac{g^2}{8} Re\arrowvert Tr(K_{\Delta}\tau^a\Delta)
+ Tr(K_{\bar{\Delta}}\tau^a\bar{\Delta})+
Tr(K_{\Omega}\tau^a\Omega)\arrowvert^2\nonumber\\
& & + \frac{g'^2}{8} Re\arrowvert Tr(K_{\Delta}\Delta)
-Tr(K_{\bar{\Delta}}\bar{\Delta})
\arrowvert^2\nonumber\\
&=&\frac{g^2}{8} Re\arrowvert 2Tr(\Delta^{\dagger}\tau^a\Delta)
+\frac{2C_L}{M_{Pl}}
Tr(\Omega\Delta^{\dagger}\tau^a\Delta)\nonumber\\
& &+2Tr(\bar{\Delta}^{\dagger}\tau^a\bar{\Delta}) +
\frac{2C_L}{M_{Pl}}Tr(\Omega\bar{\Delta}^{\dagger}\tau^a\bar{\Delta})\nonumber\\
& & + 4Tr(\Omega \tau^a\Omega)+\frac{2C_L}{M_{Pl}}[Tr(\Delta^{\dagger}
\Delta \tau^a\Omega)+Tr(\bar{\Delta}^{\dagger}\bar{\Delta}\tau^a\Omega)]\arrowvert^2\nonumber\\
& & + \frac{g'^2}{8}Re\arrowvert 2Tr(\Delta^{\dagger}
\Delta)-2Tr(\bar{\Delta}^{\dagger}\bar{\Delta})\nonumber\\
& & +\frac{2C_L}{M_{Pl}}Tr(\Delta\Omega\Delta^{\dagger})
-\frac{2C_L}{M_{Pl}}Tr(\bar{\Delta}\Omega\bar{\Delta}^
{\dagger})\arrowvert^2
\end{eqnarray}

The D-term vanishes after putting the vev's for the corresponding
fields. From above it is clear that we can not find $1/M_{Pl}$ suppressed terms
from $V_D$. So we have to go for $V_F$ to find the desired terms. 
Here we consider the first term appearing in Eq.(\ref{eqn:V_F}) i.e.
$(k^{-1})^{*k}_l W^*_k W^l$.

Substituting the Eqs.(\ref{eq:ams-nr-W}) and (\ref{eqn:AMSkahler})
in Eq.(\ref{eqn:V_F}), the terms which contribute are:
\begin{eqnarray}
 V_R &\sim&
\frac{ad_R}{2M_{Pl}}{\rm  Tr}\Omega_c^2  {\rm  Tr}\Omega_c \bar{\Delta}_c
 {\bar{\Delta}_c}^\dagger
+ \frac{a_R}{M_{Pl}} m_\Delta {\rm  Tr}\Delta_c \bar{\Delta}_c{\rm  Tr}\bar{\Delta}_c {\bar{\Delta}_c}^\dagger\nonumber\\
&+ &\frac{a_R}{M_{Pl}} m_\Delta {\rm  Tr}\Delta_c \bar{\Delta}_c
{\rm  Tr}\Delta_c\Delta_c^\dagger+\frac{ad_R}{2M_{Pl}}{\rm  Tr}\Omega_c^2
{\rm  Tr}\Omega_c\Delta_c\Delta_c^\dagger\nonumber\\
&+& \frac{c_R}{M_{Pl}}m_\Omega({\rm  Tr}\Omega_c^2)^2 + 
\frac{ac_R}{M_{Pl}} {\rm  Tr}\Omega_c^2{\rm  Tr}\Delta_c\Omega_c\bar{\Delta}_c
+ {\rm terms\,higher\,order\,in\,1/M_{Pl}}
\label{eq:ams-nr-V_R}
\end{eqnarray}
In the ABMRS model we have the relation:
\begin{equation}
 M^2_{B-L} \simeq M_R M_W; \, \omega = -|\frac{m_{\Delta}}{a}|
\equiv M_R ; \,d = \bar{d} = \left( 2 m_{\Delta} m_{\Omega}/{a^2}
\right)^{1/2} \equiv M_{B-L}
\end{equation}
After putting the vev's for the corresponding fields and
making use of appropriate scale, the terms upto $O(1/M_{Pl})$ are
\begin{equation}
 V_R \sim\frac{a(c_R + d_R)}{M_{Pl}}M_R^4M_W + \frac{a(a_R + d_R)}{M_{Pl}}
M_R^3M_W^2
\label{eq:ams-veffR}
\end{equation}

\section{BM Model}
\label{sec:ap2}
The leading order terms in this model comes from the first
term of Eq.(\ref{eqn:V_F}). writing explicitly the individual terms
\begin{equation}
 V_F = \left(K^{-1}\right)_{\Delta_c \Delta_c^{\dagger}} 
W^*_{\Delta_c} W^ {\Delta_c}+ 
\left(K^{-1}\right)_{\bar{\Delta}_c \bar{\Delta}_c^{\dagger}} 
W^*_{\bar{\Delta}_c} W^ {\bar{\Delta}_c} + \left(K^{-1}\right)_{S S^*}
W^*_S W^S
\end{equation}
Calculating the above terms the terms in the scalar potential
in lowest order in $1/M_{Pl}$ are given by,
\begin{eqnarray}
V_F &\sim&
\frac{a_R}{M_{Pl}} m_\Delta {\rm Tr}\Delta_c \bar{\Delta}_c
{\rm  Tr}\bar{\Delta}_c\bar{\Delta}_c^\dagger+
\frac{a_R}{M_{Pl}} \lambda^* S {\rm  Tr}\, \Delta_c \bar{\Delta}_c 
{\rm  Tr}\bar{\Delta}_c\bar{\Delta}_c^\dagger \nonumber \\
&+& \frac{C_R}{M_{Pl}}\left[\arrowvert \lambda \arrowvert^2 S^3
{\rm Tr} {\Delta}_c {\Delta}_c^{\dagger}+
M_{\Delta}^2 S {\rm Tr} {\Delta}_c {\Delta}_c^{\dagger}+
S^2 \left(\lambda^* M_{\Delta} {\rm Tr} {\Delta}_c {\Delta}_c^{\dagger}
+\lambda M_{\Delta} {\rm Tr} {\Delta}_c {\Delta}_c^{\dagger} \right)\right]\nonumber\\
&+& \frac{C_R}{M_{Pl}}\lambda S {\rm Tr}\Delta_c \bar{\Delta}_c
{\rm  Tr}\bar{\Delta}_c\bar{\Delta}_c^\dagger + \frac{C_R}{M_{Pl}}M_s
S^2 {\rm Tr}\Delta_c \bar{\Delta}_c + \frac{C_R}{M_{Pl}} s^3 
{\rm Tr}\Delta_c \bar{\Delta}_c\nonumber\\
 &+& {\rm Other\,terms}
\end{eqnarray}
After putting the vev's for the neutral components of the
triplet field and using the appropriate scale the term in the
highest power of $M_R$,
\begin{equation}
  V_R \sim\frac{a_R}{M_{Pl}}M_R^5 + \frac{a_R}{M_{Pl}} s M_R^4
+ \frac{C_R}{M_{Pl}} s^2 M_R^3 +  \frac{C_R}{M_{Pl}} s^3 M_R^2+
{\rm Other\,terms}
\label{eq:BMVF}
\end{equation}


\providecommand{\Di}{Di}

\end{document}